%% file: aaai22_main.tex
\def\year{2022}\relax
\documentclass[letterpaper]{article} 
\usepackage{aaai22}  
\usepackage{times}  
\usepackage{helvet}  
\usepackage{courier}  
\usepackage[hyphens]{url}  
\usepackage{graphicx} 
\usepackage{natbib}  
\usepackage{caption} 
\DeclareCaptionStyle{ruled}{labelfont=normalfont,labelsep=colon,strut=off} 
\input{dependencies}

\title{Split HE: Fast Secure Inference Combining Split Learning and Homomorphic Encryption}

\vspace{-0.3cm}

\author{
    George-Liviu Pereteanu \textsuperscript{\rm{1} \rm{2}},
    Amir Alansary \textsuperscript{\rm{1}},
    Jonathan Passerat-Palmbach \textsuperscript{\rm{1} \rm{2}}
}
\affiliations{
    \textsuperscript{\rm 1}Department of Computing, Imperial College London\\
    \textsuperscript{\rm 2}Equideum Health 

}

\begin{document}

\maketitle
\begin{abstract}
\input{abstract}
\end{abstract}

\input{body}


\input{aaai22_main.bbl}
\end{document}

%% file: dependencies.tex
\usepackage{tabularx}
\usepackage{mathtools}
\usepackage[toc,page]{appendix}

\usepackage[english]{babel}
\usepackage[utf8]{inputenc}
\usepackage[T1]{fontenc}
\usepackage{subcaption}
\usepackage{mleftright}
\usepackage{booktabs}
\usepackage{amsfonts}       
\usepackage{nicefrac}       
\usepackage{microtype}      
\usepackage{xcolor}         

\usepackage{array}
\usepackage{multirow}
\usepackage[parfill]{parskip}
\usepackage{listings}
\usepackage{color}

\definecolor{codegreen}{rgb}{0,0.6,0}
\definecolor{codegray}{rgb}{0.5,0.5,0.5}
\definecolor{codepurple}{rgb}{0.58,0,0.82}
\definecolor{backcolour}{rgb}{0.95,0.95,0.92}
 
\lstdefinestyle{mystyle}{
    backgroundcolor=\color{backcolour},   
    commentstyle=\color{codegreen},
    keywordstyle=\color{magenta},
    numberstyle=\tiny\color{codegray},
    stringstyle=\color{codepurple},
    basicstyle=\footnotesize,
    breakatwhitespace=false,         
    breaklines=true,                 
    captionpos=b,                    
    keepspaces=true,                 
    numbers=left,                    
    numbersep=5pt,                  
    showspaces=false,                
    showstringspaces=false,
    showtabs=false,                  
    tabsize=2
}
 
\usepackage{caption} 

%% file: abstract.tex
This work presents a novel protocol for fast secure inference of neural networks applied to computer vision applications. It focuses on improving the overall performance of the online execution by deploying a subset of the model weights in plaintext on the client's machine, in the fashion of SplitNNs. We evaluate our protocol on benchmark neural networks trained on the CIFAR-10 dataset using SEAL via TenSEAL and discuss runtime and security performances. Empirical security evaluation using Membership Inference and Model Extraction attacks showed that the protocol was more resilient under the same attacks than a similar approach also based on SplitNN. When compared to related work, we demonstrate improvements of 2.5x-10x for the inference time and 14x-290x in communication costs.



%% file: body.tex
\section{Introduction}

Machine learning (ML) has been widely utilised for various applications, from intrusion detection to movie recommendations. Such applications require advanced data collection techniques, which can be expensive and problematic in most cases. Privacy-Preserving Machine Learning comes up with a new way of building systems without requiring data transfer or direct access.

One fundamental approach used to directly allow the receiver to operate on encrypted data while maintaining the privacy of the systems is Homomorphic encryption (HE). Homomorphic encryption is a family of encryption schemes with a particular algebraic structure that allows computations to be performed directly on encrypted data without a decryption key. Unfortunately, the cryptographic prediction protocols presented in related works
are still unsuitable for deployment in real-world applications as they involve significant workloads to compute encrypted operations during the online execution \cite{big2019handbook}.

At a high level, all the cryptographic protocols proceed by encrypting the user's input and the service provider's neural network. Further, they use various techniques for computing different operations over encrypted data to run inference over the user's input. At the end of the execution, the client gets the final prediction, and neither party learns anything else about the other's input.

Recent works in machine learning have increased the performance of Neural Networks (NN) inference in popular applications such as image classification. A commonly cited class of applications for Convolutional NNs (CNNs) is to assist medical diagnosis. For example, a large hospital with access to a large quantity of data would like to make publicly available a model to predict the lungs medical condition. In this case, the patient wishes to classify private images using the CNN. However, a user might be reluctant to share confidential data, whereas the hospital aims to keep the model parameters private.

In Section \textit{Methodology} we introduce a novel secure inference protocol that aims to improve the time and communication cost while preserving the privacy of all parties. We consider that a server has a well-trained deep neural network and would like to expose it as a service while preserving the privacy of both the data and model. To protect the user's data, we use the CKKS \cite{Cheon_Kim_Kim_Song_2017} Fully HE (FHE) scheme to share the information and perform the computations on the cloud server. We evaluate the model's privacy through custom white box Model Extraction (MEA) and Membership Inference Attacks (MIA), which assume that a malicious client has access to samples from the training dataset. We assess the fidelity of these attacks and report that in the worst-case scenario, an attacker can reconstruct the initial model with 59\% accuracy, in Section \textit{Evaluation}.


We build upon the SplitNN \cite{vepakomma2018split} technique to enable the server to split its proprietary model into three parts. The protocol requires the middle part to be sent to the client, whereas the server keeps the other two. This design provides empirical security and privacy guarantees for both parties as shown in Sections \textit{Evaluation} and \textit{Discussion}. The protocol is generic to neural networks that support splitting. Thus, there are no requirements on the type of layers used while performing the computations.

\section{Related work}
\label{sec:related}


Secure inference using homomorphic encryption has seen much interest recently. Gazelle \cite{juvekar_gazelle_2018} is a cryptographic prediction system that came up with novel solutions of combining FHE and Multi-Party Computation (MPC) techniques to improve the inference time. At the time of creation, Gazelle was considered state of the art since it managed to perform an encrypted inference on ResNet-32 \cite{he_deep_2015} neural network within ~ 82 seconds and 560MB communication. Gazelle's use of heavy cryptography in the online phase leads to efficiency and communication overheads. These major bottlenecks were addressed in Delphi \cite{mishra2020delphi}, which builds on top of Gazelle and achieves a better performance through a better design obtained from Network Architecture Search. Delphi moves the heavy cryptographic operations over LHE ciphertexts to the preprocessing phase to reduce the online cost introduced by the high number of FHE operations. The other optimisation presented in the paper comes from the replacement of ReLU activations with polynomial (specifically, quadratic)
approximations. The authors claim an improvement of 22x for the inference time and 9x for the communication cost. CryptFlow2 \cite{ratheeCrypTFlow2Practical2Party2020}, a recent secure inference framework, further improves upon Delphi with new SMPC and HE hybrid protocols. The authors propose new protocols to evaluate DReLU activation functions and improve fixed-point arithmetic over shares to achieve better performance. CryptFlow2 runs 22x-30x faster than the prior work and shows a 10x improvement in communication cost, on par with the results we report in our experiments. While not explored in this work, the optimisations introduced by CryptFlow2 are not incompatible with our protocol and could shape a future extension.

More recently, the Concrete \cite{dolev_programmable_2021} library was released, implementing a variant of the TFHE \cite{chillotti2020tfhe} fully homomorphic encryption scheme. The core of its functionality relies on a clever use of bootstrapping to implement complex numerical functions \cite{dolev_programmable_2021}. We note that this work provides implementations for both linear and non-linear layers, facilitating new applications.

Finally, \cite{hall_syft_2021} present a similar method to the one we describe in \textit{Methodology}, but with a server splitting the initial model into two parts. The main drawback is the limitation in size of the neural networks studied. Moreover, using our custom MEA, we showed that the attacker could reconstruct the model with 95\% fidelity on the MNIST dataset.

\section{Methodology}
\label{sec:methodology}

We further describe the protocol's steps that use both an offline and online phase. The offline preprocessing phase is independent of the client's input (which regularly changes) but assumes that the server's model is static: if the server's model changes, both parties would have to re-run the preprocessing phase. Our work considers a 2-party scenario. One party, the client, has access to the private input and shares encrypted information with a model owned by a server. We assume that the server has sole ownership of the model that could have been trained against plain text data before engaging in the present protocol. As an extra measure, our protocol lets the model owner take further privacy considerations by training the initial model using Differentially-Private Stochastic Gradient Descent (DP-SGD) \cite{Abadi_Chu_Goodfellow_McMahan_Mironov_Talwar_Zhang_2016} to protect the training set if required.

\begin{figure*}[thb]
    \centering
    \includegraphics[keepaspectratio, scale=0.5]{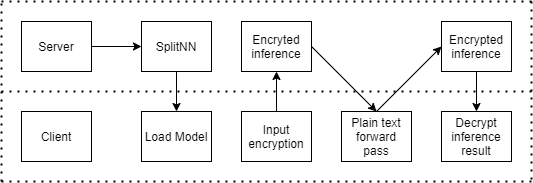}
    \caption{Encrypted inference flow of the presented protocol}
    \label{fig:steps}
\end{figure*}

Figure \ref{fig:steps} defines the information flow wherein a data subject receives the result of a prediction
performed on their private data using another entity's private model. We mention that any generic neural network can be used as long as it supports SplitNN \cite{vepakomma2018split}.

\subsection{Encryption scheme}

We're utilising the implementation of the CKKS\cite{Cheon_Kim_Kim_Song_2017} scheme as implemented in the Microsoft SEAL library\cite{sealcrypto} via the TenSEAL\cite{benaissa2021tenseal} python wrapper.

The parameters for the CKKS scheme are set as follows: we enforce 128-bits security by choosing a polynomial modulus degree of 32,768 and the scale of ciphertext up to 2$^{40}$. We additionally generate Galois keys of size 641.14MB to allow ciphertext rotations in CKKS.

\subsection{Offline phase}

\textbf{SplitNN}. The initial model is a neural network containing M layers. The server uses the SplitNN \cite{vepakomma2018split} technique to split the initial model into three models. \textit{Model 1} contains the first N layers of the initial model, \textit{Model 2} contains the next Z layers and \textit{Model 3} contains the last M - N - Z layers. In this phase, we do not require any interaction between user and server. 

\textbf{Model loading}. The server sends \textit{Model 2} in plaintext to the user, and keeps \textit{Model 1} and \textit{Model 3}. This way, we store a part of the initial model on the user's device. Deploying \textit{Model 2} to the user's device requires communication that can be done ahead of any prediction, so the time taken for sending the model is considered to be 0 in the online phase's evaluation. We consider that the user has loaded the model or can load it when necessary. Once the user receives the model, it can generate the encryption and decryption keys.

\subsection{Online phase}

\textbf{Input privacy}. The user encrypts the input homomorphically and sends it to the server at this stage. The server can now run a forward pass through \textit{Model 1} and obtains an encrypted tensor of weights to send back to the client. 

\textbf{Plain text forward pass}. To achieve a faster computation, the user decrypts the received weights to avoid encrypted-plain operations. The new input is now fed into the previously loaded plaintext \textit{Model 2}. At the end of the computation, the user will obtain a tensor of activations. This stage supports the user's data privacy since it does not involve any remote interaction.

\textbf{Encrypt the weights}. In preparation for the final stage of the protocol, the client encrypts the weights and sends them to the server. User's privacy is maintained since any information is in the form of a ciphertext.

\textbf{Final inference}. The server receives the encrypted weights and feeds them to \textit{Model 3} which will output an encrypted prediction. The server cannot decrypt the result since it is encrypted with the user's key.

\textbf{Final result decryption}. In the end, the user receives the encrypted prediction and decrypts it using their key. This phase can be considered secure since the result is encrypted during the communication. We must assume that the user kept their key private. 

\subsection{Model reduction through Distillation}

State-of-the-art neural networks achieving high accuracy on the CIFAR-10 dataset have between 25-350M parameters \cite{Krizhevsky09learningmultiple}. Because of this, the inference for a single input takes a couple of seconds.

The chosen model architecture for our study is Deep Layer Aggregation (DLA) \cite{yu_deep_2019}, with 15.3M trainable parameters. It represents the perfect balance between size and performance, achieving an accuracy of over 90\% on the dataset. Bigger models used to classify the CIFAR-10 dataset achieve 99.42\% accuracy, but the number of parameters increases up to 307M \cite{dosovitskiy_image_2021}. At the other end of the spectrum, models at the core of related works (Section \textit{Related Work}) present a smaller accuracy of around 80\% \cite{he_deep_2015} and have only 0.46M. 

To speed up the computation, we applied the principle of distilling the knowledge in a neural network \cite{hinton2015distilling}. This idea consists in creating a teacher model and storing its logits (not predictions). We create a student model with fewer parameters and train it on the logits of the original model. Distillation on DLA reduced the number of parameters from 15M to 700K and achieved an accuracy of 90.22\% for the teacher model and 90.1\% for the student model. We use the student model as the primary reference from this point onward. This model can perform this optimisation after initially training the model before the offline phase.

\section{Evaluation}
\label{sec:experiments}

\subsection{Runtime and communication}

We need to consider the number of layers each model consists of at each point. The TenSEAL framework currently experiences limitations in the number of convolutional layers that can be calculated sequentially in the encrypted realm. As a result, we split our DLA model with \textit{Model 1} holding the first (convolutional) layer, \textit{Model 2} containing the next 12 to 15 layers  and \textit{Model 3} holding the last 4 to 1 layers.

We evaluate the runtime for our protocol and report metrics for the following stages: 
\begin{enumerate}
    \item T1: user encrypts the input image
    \item T2: server runs the encrypted convolution
    \item T3: user runs the forward pass in plaintext and encrypts the result
    \item T4: server computes forward pass to get the final encrypted tensor of results
    \item C4: server sends encrypted vector to the user - negligible due to size
\end{enumerate}

C1, C2 and C3 represent the amount of communication measured in MB between client and server. Table \ref{table:protocol2-time} shows how the number of encrypted layers influences the inference time.

\begin{table*}[]
\caption{Runtime and communication evaluation based on the number of encrypted layers} 
\centering 
\begin{tabular}{|c|c|c|c|c|c|c|c|c|c|}
\hline
Encrypted layers & T1(s)                    & C1(MB)                   & T2(s)                     & C2(MB)                   & T3(s)    & C3(MB)  & T4(s)    & C4(MB)                 & Total time(s) \\
\textit{(Model 1)} & & & & & & & & & \\ \hline
1                & \multirow{4}{*}{0.05} & \multirow{4}{*}{1.7} & \multirow{4}{*}{0.056} & \multirow{4}{*}{2.6} & 0.078 & 0.1 & 0.936 & \multirow{4}{*}{0} & 1.12       \\ \cline{1-1} \cline{6-8} \cline{10-10} 
2                &                       &                      &                        &                      & 0.065 & 0.4 & 4.6   &                    & 4.771      \\ \cline{1-1} \cline{6-8} \cline{10-10} 
3                &                       &                      &                        &                      & 0.058 & 0.8 & 11.1  &                    & 11.264     \\ \cline{1-1} \cline{6-8} \cline{10-10} 
4                &                       &                      &                        &                      & 0.051 & 1.1 & 28.4  &                    & 28.557     \\ \hline
\end{tabular}
\label{table:protocol2-time} 
\end{table*}

\subsection{Comparison with State-of-the-Art}

Our second evaluation benchmarks our protocol with the works presented in Section \textit{Related Work}. We assess the different implementations by reporting the inference time, communication cost and accuracy in Table \ref{tab:Comparison table}. All protocols operate under a semi-honest setting.

 
We observe that the usual accuracy of related work on the same dataset is around 80\%. As mentioned before, our chosen architecture is different and raises the accuracy of our model to 90\%. As a result, we also report metrics for our protocol when run with the model used in the literature \cite{he_deep_2015}. Both runtime and communication costs outperform the other protocols under this scenario. Our cryptographic protocol requires only 4.4MB in the online phase when communications for the same secure inference task see Delphi using 14x and Gazelle 290x more bandwidth. The techniques presented in this work enable us to leverage more realistic and larger architectures, and as such, to obtain better accuracy on the task. In terms of inference time, we improve upon Gazelle \cite{juvekar_gazelle_2018} by 3x, Concrete \cite{dolev_programmable_2021} by 5x and Falcon by 2.5x.

Finally, we managed to run Delphi \cite{mishra2020delphi} on the same machine used for our experiments. We find that Delphi's runtime increases from 3.7s to 47.3s, making our protocol an order of magnitude faster when running on our machine.

\begin{table*}[!htb]

            \centering
            \caption{Comparison table to related work. Delphi \cite{mishra2020delphi} could be re-run on our machine. All other reported values are directly extracted from the original papers, when available. As a reference, we run the experiments on a machine with Intel Xeon Processor (Skylake) Model 85 (8 cores, 2.1 GHz), 16GB RAM and Tesla T4 GPU.}
            \begin{tabular}{|c|c|c|c|c|}
            \hline
            Protocol & Inf. time (s)                   & Comms. (MB) & Accuracy & Dataset  \\ \hline
            Ours + \cite{yu_deep_2019}                 & 4.76                                 & 4.3                & 90.1\%   & CIFAR-10 \\ \cline{1-5}
            Ours + \cite{he_deep_2015}                 & 3.6                                  & 4                  & 81.61\%  & CIFAR-10 \\ \cline{1-5}
            Delphi   & 3.8 / \textbf{47.3}                  & 60             & 81.61\%  & CIFAR-10 \\ \cline{1-5}
            Gazelle  & 12.9                                 & 1236               & 81.61\%  & CIFAR-10 \\ \cline{1-5}
            Falcon   & 10.1                                 & 1459               & 81.61\%  & CIFAR-10 \\ \cline{1-5}
            EVA      & 4.47                                 & N/A                & 79.38\%  & CIFAR-10 \\ \cline{1-5}
            Concrete & 17.96                                & N/A                & 97\%     & MNIST    \\ \hline
            \end{tabular}
            \label{tab:Comparison table}

\end{table*}

\begin{table}[htb]
    \centering
    \caption{MEA fidelity with the numbers of training samples owned by the client and encrypted layers kept at the server performed on CIFAR-10 dataset}
    \begin{tabular}{|c|c|c|}
    \hline
    \# samples & \# encrypted layers & Attack accuracy \\
    (share of dataset) & in \textit{Model 3} &  \\ \hline
    500        & 4                   & 38.14\%         \\ \hline
    500        & 3                   & 43.24\%         \\ \hline
    500        & 2                   & 48.63\%         \\ \hline
    500        & 1                   & 50.8\%          \\ \hline
    2500       & 4                   & 41.12\%         \\ \hline
    2500       & 3                   & 50.76\%         \\ \hline
    2500       & 2                   & 54.17\%         \\ \hline
    2500       & 1                   & 56.38\%         \\ \hline
    5000       & 4                   & 42.19\%         \\ \hline
    5000       & 3                   & 53.02\%         \\ \hline
    5000       & 2                   & 57.56\%         \\ \hline
    5000       & 1                   & 59.4\%          \\ \hline
    \end{tabular}
    \label{tab:mscenario2_attack}
\end{table}

\subsection{Security evaluation}

As Section \textit{Methodology} explained, the protocol limits the trust boundaries and allows a malicious client to access a significant part of the trained model. The utility of our proposed solution is shown so far by the improvement over the inference time, reduction in communications and ability to handle large models. In a real-world scenario, a malicious user could try to leverage the plaintext weights from \textit{Model 2} to extract potentially valuable intellectual property from the model or information about the training set. We will address these two concerns with results against Model Extraction Attacks (MEA) and Membership Inference Attacks (MIA). However, the literature lacks the white-box attacks that would take advantage of our current setup. Works such as \cite{carlini2020cryptanalytic}, \cite{Truong_2021_CVPR} do not take full advantage of the plaintext weights and rely either on the weakness of the activation layers or data-free knowledge. 


To highlight the empirical robustness of our protocol against MEA, we create a scenario where 
the client only has access to the central part of the model, whereas the server controls the first and last layers. A malicious client could try to break the server's privacy and retrieve the parameters of the original model. Moreover, we assume a malicious client might also have access to samples from the training dataset to strengthen his attack. In our case, the client attacker knows the output shape of \textit{Model 1} (although kept secret on the server's side) since it will serve as input to \textit{Model 2}. For example, an attacker sending an input image of size 32x32 will receive a matrix of the same size, thus revealing that \textit{Model 1} does not modify the dimensions.

The attacker creates two new models to reconstruct the missing parts corresponding to \textit{Model 1} and \textit{3}. We studied both cases where the attacker knows the original architecture or not. In the end, to reconstruct the missing parts, we observed that the fidelity grows when the attacker utilises a new CNN architecture with five convolutions. The first missing part is trained by minimising the difference between the model output and server response to \textit{Model 1}. The last part of the missing model aims to reproduce the last layers up to the final inference. The same procedure is also applied to the second trainable model, with the targets now set to the actual known values.  Table \ref{tab:mscenario2_attack} shows the fidelity of the reconstructed model. We report multiple cases where the server hides up to the last four layers. In the worst-case scenario, when the server does not try to protect the model (i.e. it keeps only the last layer), an attacker can reconstruct the model with 59\% fidelity.

For the sake of comparison, we attacked the protocol in \cite{hall_syft_2021} using the same MEA on MNIST. The final results show that an attacker who has access to only 1\% of the samples from the original training dataset could reconstruct the initial model with 98\% fidelity. Under the same conditions, our protocol would only allow an attacker to reconstruct with 38.14-50.8\% fidelity.


To determine the security of our protocol from a data owner's perspective, we conducted a new experiment based on a Membership Inference Attack (MIA). The MIA scenario involves an attacker aiming to learn whether a targeted individual's data was used while training a model. Our implementation relies on the \texttt{mia} library \cite{bogdan_kulynych_2018_1433745}.

The attack requires training several shadow models with the same architecture as the target model on different datasets sampled from the original data distribution. In the literature, Shokri et al \cite{shokri2017membership} suggested a representative attacker might have access to 10\% of samples from the victim's dataset. Running attacks with the pre-set values for training size (10\%) and shadow models (3), a malicious client could estimate the presence of an individual's data point in the training set with \textbf{52.3\%} precision. In this case, an adversary makes a random guess about the presence of an individual in the dataset.

We further assess our protocol by increasing the attacker's knowledge about the victim's data distribution to less realistic settings. We used 20\% samples from the original CIFAR-10 dataset and 20 shadow models. The results have shown a slight increase in the accuracy, which increases to \textbf{59.6\%}. Here the attacker obtains a slightly better prediction than a random guess in what appears as an overly conservative setting from what could be described as a worst-case scenario.

We also implemented the first MIA attack using Tensorflow Privacy. To run the best possible experiment using this library, we consider the two types of attacks available: Threshold attacks, inspired by \cite{song2021systematic}, which report an AUC score, and Trained attacks (K nearest neighbours, Logistic regression and Multi-layered perceptron), based on \cite{yeom2018privacy} that return the adversary's advantage. Table \ref{table:nonlin} presents the experiment's scores on a specific class and the entire dataset, displaying only the highest entries for AUC and Advantage scores across a  class. While Tensorflow Privacy's documentation states that scores from different types of attacks are not directly comparable, we observe that the AUC scores establishing the capability an attacker has to distinguish between member and non-member are similar to our \textit{mia} based implementation. Moreover, the attacker's advantage remains low and consistent across all classes. This observation concurs with the previous results that the model does not leak sufficient information to do better than a random guess.

\begin{center}
\begin{table*}[htb]
\caption{MIA attack on MNIST using Tensorflow Privacy - Missing entries denote that another attack bested the one that would have appeared in the empty column.
} 
\centering 
\begin{tabular}{c c c c c} 
\hline 
\textbf{Class} & \textbf{Threshold} & & \textbf{Trained Attacks} & \\ 
 & \textbf{Attacks} & K nearest & Logistic regression & MLP \\ [0.5ex] 
\hline 
Class 0 & & AUC - 0.52 & & Adv - 0.06 \\
Class 1 & & & AUC - 0.54, Adv - 0.11 &  \\
Class 2 & & & AUC - 0.51, Adv - 0.06 & \\
Class 3 & & & AUC - 0.52, Adv - 0.1 & \\
Class 4 & & & AUC - 0.51 & Adv - 0.06 \\
Class 5 & & & &AUC-0.52, Adv-0.1 \\
Class 6 & & & AUC - 0.52, Adv - 0.09 & \\
Class 7 & & AUC - 0.52 & Adv - 0.07 & \\
Class 8 & & AUC - 0.54 & & Adv - 0.1 \\
Class 9 & & & AUC - 0.55 & Adv - 0.1 \\
Entire dataset & & & Auc - 0.51, Adv - 0.03 & 
\\ [1ex] 
\hline 
\end{tabular}
\label{table:nonlin}
\end{table*}
\end{center}

\section{Discussion}
\label{sec:discussion}

The initial set of experiments conducted in this work allow us to draw some conclusions and open exciting perspectives for further studying the approach and improving upon this first version. In our current setting using the DLA model, we observe from Table \ref{tab:mscenario2_attack} that encrypting four layers in Model 3, up to the fourth layer from the right end of the network, has a significant negative impact on the MEA's fidelity. At the same time, Table \ref{table:protocol2-time} reveals that the addition of the fourth encrypted layer also has a disproportionate impact on the server's processing time. We attribute this lower performance in part to the underlying framework, Tenseal, and the machine we have run our experiments on. Indeed, the latest release of SEAL, which powers Tenseal, can leverage modern hardware acceleration available on cutting-edge Intel CPUs \cite{intel_hexl}. The results we report should thus be considered in light of the potential improvement that would come naturally by reimplementing our protocol on a better-optimised hardware and software stack. Switching to another HE stack would also allow us to experiment with the number of encrypted layers at both ends of the model.

When it comes to Security and Privacy, the main limitation of our protocol is that it does not provide any formal privacy guarantee, as it suffers from the same downsides as Split Learning. SL's security has been challenged since its introduction, and attacks are proposed regularly \cite{unleash_tigger}. At the same time, improvements the base protocol regularly receives improvements that primarily aim to reduce the likelihood for smashed data to be leveraged as part of a reverse engineering attempt \cite{vepakomma2020nopeek, singh2021disco}. While our protocol's initial version builds upon the original SL algorithm for simplicity, future iterations will incorporate the latest updates to SL.
Additionally, a cautious data owner's privacy could train the model deployed with our protocol using Differential Privacy (DP). DP would provide formal upper bounds of the privacy leakage induced by sharing the model in plaintext. The upper bound would likely not be tight since our protocol does not provide full white-box access to potential attackers.

\section{Conclusion}

In this paper, we have introduced a new protocol for fast secure inference of neural networks. The design is generic to all networks that support splitting \cite{vepakomma2018split} in three parts, in order to drastically reduce the amount of encrypted computations. To ensure both the privacy of the initial model and the user's data, our protocol hides the most sensitive layers of the NN on the server.

Our approach delivers state-of-the-art runtime performance and can also leverage larger networks than other works in the literature. We have empirically assessed the security of the protocol by attacking it under MEA and MIA settings, showing stronger resistance than a similar approach in \cite{hall_syft_2021}.


Future work will focus on improving the performance of our implementation and testing it against different FHE libraries. From a security standpoint, we aim to asses the viability of our protocol against more complex attacks that could better leverage the white-box setting. We will also explore more models and datasets to confirm the patterns observed when conducting empirical attacks.